# Bell's inequality and Quantum Probability Trees

Hamidreza Simchi *

Semiconductor Component Industry , P.O.Box : 19575-199 , Tehran - Iran

March 5, 2002

**Abstract**

The relations between Bell's inequality and quantum probability trees are explained against the background offered by the concept of a quantum probability tree built in others works [1] . It is shown , that , if we use the concept of probability tree , it will not be necessary , we set aside the principle of separability and principle of locality .

## 1   introduction

The famous paper about complete description of pysical reality , which , was written by Einstein , Podolsky and Rosen (EPR-paradox)(2) , includes , the following important subjects[3] :

1. There is physical reality , independent from our knowledge.

2. The aim of science is the discovery of this reality .

3. A scientific theory will not be complete, if it does not include the physical reality .

4. The necessary condition for a complete physical theory is, that , every element of physical reality has a representation in theory .

5. The sufficient condition for assigning a quantity to physical reality is , that , we can measure it without interference with system .

Also Einstein , in another paper [4] , reminded , that , for escaping from lack of quantum mechanic , we must set aside at least one of two important principles , which , are principle of separability and principle of locality [3] . Based on these principles and important subjects , a formal spin-correlation measurement was done

---

*e-mail: srdc@isiran.com

. In this measurment , a two-electron system in a spin-singlet state was considered .
If $a$ , $b$ and $c$ are considered as unit vectors , which , are not mutually orthognal , it
was shown that the following inequality holds [5] :

$$P(a+,b+) \leq P(a+,c+) + P(c+,b+) \tag{1}$$

Which , $P(a+,b+)$ is the probability that , in a random selection , for particle one $(S_1.a)$ to be $(+)$ and $(S_2.b)$ to be $(+)$ and so on . This is called Bell's inequality . Here , after a brief review on the concept of quantum probability tree [1] , I explain the spin-correlation experiment based on the concept of quantum probability trees , and , show if we use the concept of information probability and note to the concept of probability tree it is not necessary , one of the two principles i.e. principle of separability and principle of locality is set aside .

## 2 Quantum probability tree [1] :

The key concept for finding probability tree is : " A quantum mechanical random phenomenon consists of a sequence of two distinct operations , an operation $P_\Psi$ of state preparation and a measurement operation $M_\Omega$ that ends with the registration of a " needle position " $V_{\Omega j}$ of a macroscopic device $D_\Omega$ for measurement of the observable $\Omega$ , whose , eigenvalues are $w_j$. "
Based on this key concept , we can define two probability chains as follows : Formal probability Chains :

$$[(\Psi, \Omega), \{w_j\}] \longrightarrow [\{w_j\}, \tau, \Pi(\Psi, \Omega)] \qquad (2)$$

Factual probability Chains :

$$[(P_\Psi, M_\Omega), \{V_{\Omega j}\}] \rightarrow [V_{\{\Omega j\}}, \tau_F, P(P_\Psi, M_\Omega)] \qquad (3)$$

Which :
$\tau \equiv$ the total algebra on $\{\omega_j\}$
$\Pi(\Psi, \Omega) \equiv$ the probability density law on $\tau$.
$\tau_F \equiv$ the total algebra on $\{V_{\Omega j}\}$
$P(P_\Psi, M_\Omega) \equiv$ the probability measure on $\tau_F$

These two chains are connected as :

$$P(P_\Psi, M_\Omega, V_{\Omega j}) \Rightarrow \Pi(\Psi, \Omega, \omega_j) = |<u_j|\Psi>|^2 \qquad (4)$$

where:

$\Omega|uj> = \omega_j|u_j>$.

Of course two distinct measurement ( classes )$M_x$ and $M_y$ are mutually incompatible , If it is not possible to construct physically an individual measurement evolution such that the unique outcome$V_{xj}$ produced by it permits calculation of both a corresponding eigenvalue tied with $M_x$ and a corresponding eigenvalue tied with $M_y$,
( Bohr Complementarity ) , otherwise they are compatible .
The situation entails that , globally , the whole , the unity constituted by the ensemble of all the factual probability Chains corresponding to a fixed operation of state preparation $P_\Psi$ possesses a branching , a tree like space-time structure .
It is symbolized by $T(P_\Psi)$ and is called, " The quantum mechanical probability tree of the operation of state preparation$P_\Psi$". From information probability point of view , quantum probability can be written as :

$P_\Psi \equiv$ information source with zero memory , such that , unique outcome $|\Psi>$ is produced by it .

$M_x, M_y, ... \equiv$ quantum information channel , such that , after connecting to information source, outcomes $V_{xj}, V_{yk}, ...$ are produced by it . They are shown by $C_x, C_y, ...$ .

$I(P_\Psi, C_x), I(P_\Psi, C_y), ... \equiv$ nondeterministic information systems without noise corresponding to a set of observable $\Omega_x^h, h = 1, 2, ..., l$ all tied to the same class $M_x$ of individual measurement evolution , or , respectively , to a set of observable $\Omega_y^g = 1, 2, ..., s$ all tied to another same class $M_y$ of individual measurement evolution , etc .

# 3 Spin-Correlation Measurement

By an operation $P_\Psi$ of state preparation , a two electrons system in a spin-singlet state is prepared . $M_\Omega$ is stand for spin measurement in three different directions $a$, $b$ , $c$ , which , are unit vectors ( not mutually orthogonal ) and $V_{\Omega j}$ is the position of pointer of measurement tools , which , can be up (+) , or , down (-) .

Also $M_{12}^a$ is measurement in a direction , $M_{12}^b$ , in b direction and $M_{12}^c$ in c direction . Since spin operator of first particle in a direction ,$\Omega_1^a$ commute with spin operator of second particle in a direction , $\Omega_2^a$ by single measurement $M_{12}^a$ , we can find both of them and therefore , this measurement is called $M_{12}^a$ , which , shows by single measurement two eigenvalues ( one of first and other of second particle ) are found simultaneously . In the otherwords , by an operation $P_\Psi$ , we make ( prepare ) a complex system , with space-time domain $\Delta(P_\Psi) = \Delta x \Delta t$, which , will be trunk of tree .

Then in the case of an evolution $M_{12}^a$ corresponding to the two commuting observable $\Omega_1^a$ and $\Omega_2^a$ , the subsequent phase of measurement covers , for each fiber corresponding to the process $M_{12}^a$ of measurement evolution, a unique space-time domain $\Delta(M_{12}^a)$. The measurement $M_1^a$ and $M_2^a$ are meaningless here . If we had a operation $P_{\Psi_1}$ and another operation $P_{\Psi_2}$ they can be considered the trunks of two different trees with two domains $\Delta(P_{\Psi 1}) = \Delta x_1 \Delta t_1$ and $\Delta(P_{\Psi 2}) = \Delta x_2 \Delta t_2$ and then it will be possible , that , fibers $M_1^a$ with $\Delta(M_1^a)$ and $M_2^a$ with $\Delta(M_2^a)$ are defined ( Considered) .

In the other words , by operation $P_\Psi$ , we encounter with a ocean of many potentials , which , is called information system S but each potential includes two electron in a spin-singlet state , $|\Psi>$, and not one electron $|\Psi 1>$ and other electron $|\Psi 2>$ . After connecting the channel$M_{12}^a$ to information system S , out come includes a pair of information i.e. $(a_1^+, a_2^-)$ or $(a_1^-, a_2^+)$ .

Information system , $S$ , including Spin - singlet state $|\Psi>$ , is not the sum of information system $S_1$ including many potentials related to first particle and information system $S_2$ including potential related to second particle . $M_{12}^a$ can be only defined ( act ) on $S$ and $M_1^a$ and $M_2^a$ can be only defined ( act ) on $S_1$ and $S_2$ respectively .

By this definition of quantum probability , and by operation $P_\Psi$ always we have potential of correlated particles and never we have single particles . It is not important how long . They are correlated to each other for all times . Now , the importance of the operation , $P_\Psi$ is specified . It is the first step in defining the quantum probability . Therefore , by this definition of quantum probability it is not necessary we set aside the principle of separability and principle of locality .

If we start with Kolmogorov's concept of probability , for escaping the lack of quantum mechanic we must set aside one of the these two principles , because , Kolmogorov's concept does not include the concept of operation $P_\Psi$ , random phenomena and ocean of potentials in information source $S$, and does not note to their importance in defining the probability . In Kolmogorov's concept , at first we assume the exact and specific entities exist really , not as many potentials , and second assume they are separate entities and since quantum mechanical predictions are not compatible with this inequality we conclude that we must set aside one of the two important principles . But important hint is that , quantum probability is some kind of information probability and not Kolmogorov probability [1] .

The root of this confusion and paradox is that we assumed quantum probability is Kolmogorov probability and naturally we reach to non correct result . Based on information probability , consider an information source , $S$ , with zero memory , which emits an input $(a_1^+, a_2^-)$ with input probability law $P(a_1^+, a_2^-)$ on it .

After association of channel $C$ to input source $S$ , an output $(b_1^+, b_2^-)$ with an output probability law $P(b_1^+, b_2^-)$ is found .

The total probability $P(b_1^-, b_2^+)$ is calculated as :

$$P(b_1^-, b_2^+) = \sum P(a_1^+, a_2^-) P[(b_1^-, b_2^+)/(a_1^+, a_2^-)] \tag{5}$$

Where

$$M(C/S) = P[(b_1^-, b_2^+)/(a_1^+, a_2^-)] \tag{6}$$

is certain conditional probability i.e. in the most general case any input $(a_1^+, a_2^-)$ can produce any output $(b_1^-, b_2^+)$ . $M(C/S)$ is called channel matrix of this experiment . If now , we compare (1) with (2) , we can see there are big differences between them , since , we have one state $|\Psi>$ and not $(a|\Psi_1> + b|\Psi_2>)$ .

Of course , we believe the superposition principle , when , we want to do mathematical calculations , but , the operation $P_\Psi$ prepare $|\Psi>$ , which , is mixed as one entity , and not as a mixture of $|\Psi_1>$ and $|\Psi_2>$ with separability property , such that , after experiment this mixture $|\Psi>$ breaks to two different states $|\Psi_1>$ and $|\Psi_2>$ i.e. $|\Psi>$ is mixture of $|\Psi_1>$ and $|\Psi_2>$ , but after operation $P_\Psi$ it has not separability property and therefore only $M_{12}^a$ , $M_{12}^b$ and $M_{12}^c$ can operate on it .

# 4  Conclusion

I summarize the results as follows :

1)-The quantum mechanical probability is some kind of information probability and not Kolmogorov probability .

2)-The state operation $P_\Psi$ , is very important step for calculating quantum probability and is the trunk of probability tree .

3)-By operation $P_\Psi$ mixed and complex spin singlet state are formed as ocean of potentials and source of information , $S$ . It is not the sum of information source $S_1$ and information source $S_2$ . Superposition principle is valid , but after operation $P_\Psi$ , mixed state $|\Psi>$ exists and can not be broken to $|\Psi_1>$ and $|\Psi_2>$ .

4)-The information channels are $M^a_{12}$ , $M^b_{12}$ and $M^c_{12}$ which can be connected to source of information , S. Channels $M^a_1$ , $M^a_2$ and so on can only connect to source $S_1$ and $S_2$ respectively and not $S$ .

5)-Since we used non correct concept i.e. Kolmogorov's concept of probability instead of information probability we become forced for setting aside the principle of separability and principle of locality . If we use information probability concept , since we have mixed and complex state $|\Psi>$ for all times , which , is prepared by operation $P_\Psi$ as a trunk of probability tree , it is not necessary we set aside these principles .

6)-Of course , we can assume the ocean of potential of state really exist as a reality [ but complex state $|\Psi>$ , not single states $|\Psi_1>$ and $|\Psi_2>$] but finality does not exist . i.e. we can not specify in which state the system is placed now else we do measurement M. Therefore before measurement an ocean of reality exist and finality does not exist but after doing measurement $M$ , not only ocean of reality exist but also finality exist i.e. we know that the system is in which state now .